# Stabilization of tetragonal/cubic phase in Fe doped Zirconia grown by atomic layer deposition

A. Lamperti [a),*], E. Cianci [a)], R. Ciprian [a)], D. Sangalli [a)], and A. Debernardi [a)]

[a)] *Laboratorio MDM, IMM-CNR, via C. Olivetti 2, 20864 Agrate Brianza (MB), Italy*

**Abstract**

Achieving high temperature ferromagnetism by doping transition metals thin films is seen as a viable approach to integrate spin-based elements in innovative spintronic devices. In this work we investigated the effect of Fe doping on structural properties of $ZrO_2$ grown by atomic layer deposition (ALD) using $Zr(TMHD)_4$ for Zr and $Fe(TMHD)_3$ for Fe precursors and ozone as oxygen source. The temperature during the growth process was fixed at 350°C. The ALD process was tuned to obtain Fe-doped $ZrO_2$ films with uniform chemical composition, as seen by time of flight secondary ion mass spectrometry. The control of Fe content was effectively reached, by controlling the ALD precursor pulse ratio, as checked by X-ray photoemission spectroscopy (XPS) and spectroscopic ellipsometry. From XPS, Fe was found in $Fe^{3+}$ chemical state, which maximizes the magnetization per atom. We also found, by grazing incidence X-ray diffraction, that the inclusion of Fe impurities in $ZrO_2$ induces amorphization in thin $ZrO_2$ films, while stabilizes the high temperature crystalline tetragonal/cubic phase after rapid thermal annealing at 600°C.

* corresponding author: Alessio Lamperti, Laboratorio MDM, IMM-CNR, Via C. Olivetti 2, 20864 Agrate Brianza (MB), ITALY (email:alessio.lamperti@mdm.imm.cnr.it).

## 1. Introduction

The capability to control both charge and spin in a material by using a magnetic or electric field at room temperature is a key factor in qualifying that material as potential candidate to be integrated in spin-based elements of innovative spintronic devices. A viable route to achieve high temperature ferromagnetism consists in engineering, for instance by doping, transition metal oxides or semiconductors. Based on the theoretical prediction that Mn doping in ZnO would generate magnetism mediated by the insertion of the dopant [1], many experiments have been conducted, in the last years, on different oxides, such as $TiO_2$, $ZnO_2$, $SnO_2$, $HfO_2$ [2] and, recently on Mn doped monoclinic $ZrO_2$ [3]. From *ab-initio* calculations Mn doped $ZrO_2$ has been predicted to exhibit ferromagnetism with Curie temperature above 500°C when doped with Mn concentration up to 40% [4] and experimentally confirmed at 400 K for Mn concentration of 5% [3]. Despite doping allows transition metals to gain ferromagnetism, its effective origin has been shown to crucially depend on the role played by the defects, mainly oxygen vacancies [5,6,7,8,9,10], which are indirectly introduced during the doping process. If the observed ferromagnetism were related to the presence of defects, in principle, it would be very difficult to control and integrate these materials in foreseen applications, unless the stability of the defects will be much improved to sustain, for instance, the thermal budget experienced in a standard fabrication process, which typically requires exposure at 300-600°C temperature.

Among the methods and processes proposed for the deposition of pure and doped $ZrO_2$ thin films such as sputtering, chemical vapor deposition or pulse laser deposition, atomic layer deposition (ALD) is particularly advantageous for the growth of very thin films because, being a self-limiting process, it is possible to carefully control the film thickness, uniformity and conformality, even over large areas, while maintaining a relatively low thermal budget during the film growth [11]. A broad literature reports the successful ALD of pure and doped $ZrO_2$, with different types of Zr precursors (see for instance [12] and references therein) and doping elements, such as Al [13], Er [14] and La [15,16]. In particular, ALD doped $ZrO_2$ thin films stabilize the cubic or tetragonal crystalline $ZrO_2$ phases, which is also associated with the presence of defects, such as oxygen vacancies.

Based on the above considerations, we developed and optimized an ALD process for the growth of $ZrO_2$ thin films doped with Fe, which, in principle, should be beneficial both in terms of favoring $ZrO_2$ cubic/tetragonal phase (i.e. vacancy formation) and possibly act in promoting ferromagnetism, similarly to Mn doping in monoclinic $ZrO_2$. We found that Fe induces amorphization in as grown thin $ZrO_2$ films, while stabilizes the high temperature crystalline tetragonal/cubic phase with respect to the monoclinic phase in pure zirconia after rapid thermal annealing above 600°C. Our experimental evidence is in agreement with density functional theory (DFT) based simulations, predicting that Fe doping induces a crystalline phase transition and the formation of O vacancies, which ultimately, at significant concentration, makes the material behave as a ferromagnetic semiconductor.

## 2. Experimental details

Pure and Fe doped $ZrO_2$ thin films were grown on $Si/SiO_2$ substrates in a flow-type hot wall ALD reactor (ASM F120) starting from β-diketonates metalorganic precursors, namely $Zr(TMHD)_4$ for Zr and $Fe(TMHD)_3$ for Fe (TMHD=2,2,6,6-tetramethyl-3,5-heptanedionate). To grant a stable reactivity, Zr precursor was kept at 170°C, while Fe precursor was maintained at 115°C. Ozone was used as oxidizing gas in the reaction process. The film growth was achieved by alternately introducing the pulses of the reactants separated by $N_2$ inert gas purging pulses, as schematically sketched in Figure 1 for the case of Fe doped $ZrO_2$ ($ZrO_2$:Fe) film deposition. The growth temperature was maintained at 350°C. After deposition, films were annealed at 600°C or 800°C in $N_2$ flux for 60s to study film thermal stability.

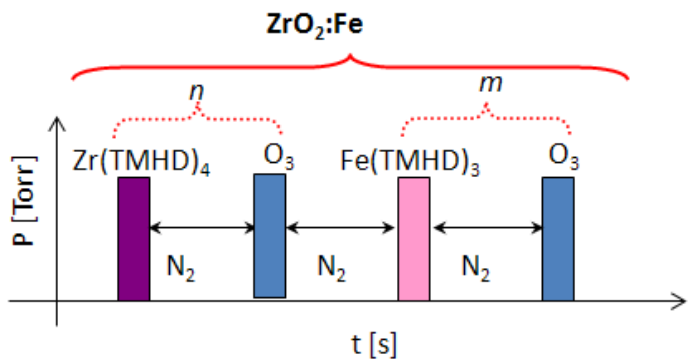

**Figure 1** Schematic of the ALD process used for the growth of $ZrO_2$:Fe films. Alternate pulses of metallic (Zr or Fe) or $O_3$ precursors are separated by purging pulses using $N_2$ gas.

Film thickness was monitored by spectroscopic ellipsometry (SE) (Woollam M-2000 F) and further calculated, along with surface roughness and electron density by specular X-ray reflectivity (XRR) (Italstructure XRD 3000). Film crystallinity was checked by X-ray diffraction (XRD) at fixed grazing incidence angle ω = 1° and using Cu Kα (wavelength = 0.154 nm) monochromated and collimated X-ray beam. Further details on the configuration and parameters used in performing X-ray measurements can be found in [12].

Film chemical composition and uniformity was checked by time-of-flight secondary ion mass spectrometry (ToF-SIMS) depth profiles (ION TOF IV), using 500 eV $Cs^+$ ions for sputtering over 200 × 200 μm$^2$ area and 25 keV $Ga^+$ ions for the analysis of 50 × 50 μm$^2$ area centered on the sputtered area, in negative polarity and interlaced mode [17]. Secondary ions intensity has been normalized to $^{30}Si$ intensity in bulk Si substrate. To elucidate Fe chemical state and concentration in $ZrO_2$:Fe films grown with different $m:n$ = Fe:Zr binary process pulse ratio, dedicated XPS measurements were performed on a PHI 5600 instrument equipped with a monochromatic Al Kα x-ray source (energy = 1486.6 eV) and a concentric hemispherical analyzer. The spectra were collected at a take-off angle of 45° and band-pass energy 23.50 eV. The instrument resolution is 0.5 eV.

## 3. Results and discussion

The growth process was calibrated by growing films in the thickness range 10-20 nm and considering the number of ALD cycles employed to complete each growth. Figure 2 shows that film thickness, measured by SE, linearly increases as a function of the ALD cycles, providing strong indication that the deposition of $ZrO_2$ film is self-limited and within the so-called ALD window regime.

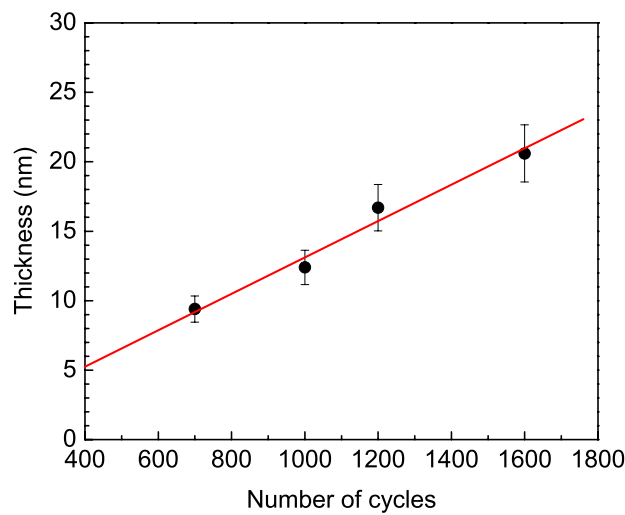

**Figure 2** – Linear relation of the film thickness with number of cycles indicates that films grow in ALD regime at a growth rate of 0.13 Å/cycle.

From the slope of the linear fit a growth rate of 0.13 Å/cycle is found, when pure $ZrO_2$ films are considered. It is worth noticing that such a low growth rate could be advantageous, when Fe doping is also inserted, as we can control the amount of Fe also in very thin films.

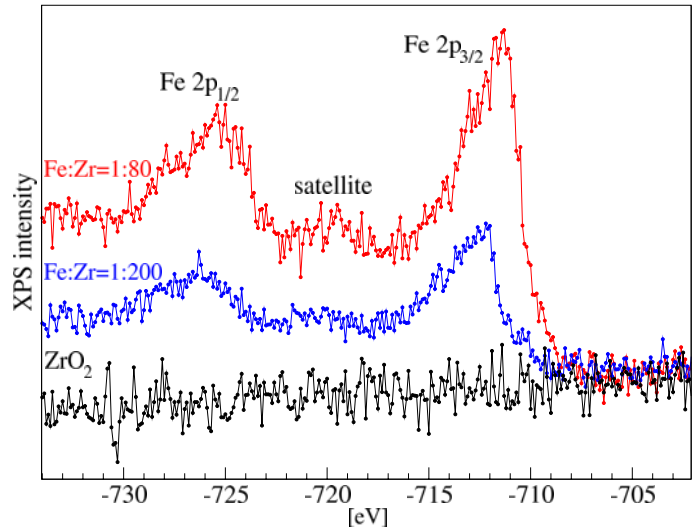

**Figure 3** – XPS measurements of $ZrO_2$ and Fe doped $ZrO_2$ films at different Fe doping.

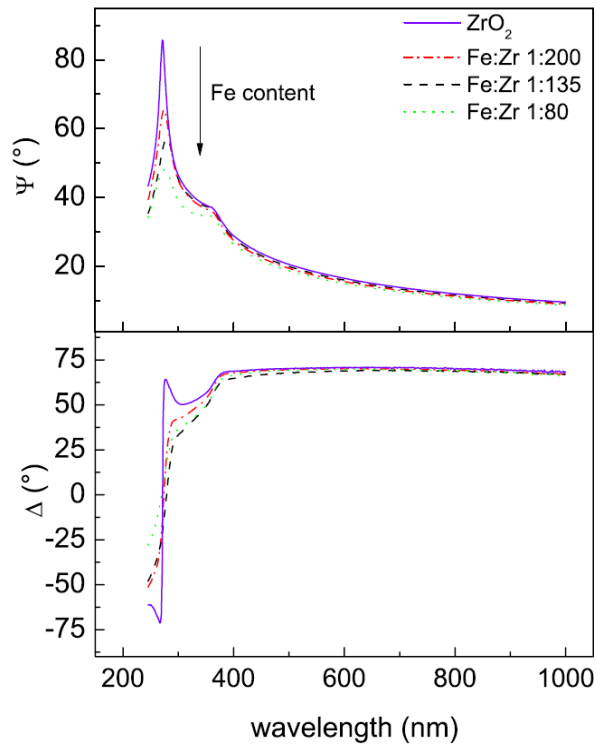

**Figure 4** - SE experimental and model fit data of $ZrO_2$ and Fe doped $ZrO_2$ films at different Fe doping. The variation in $\Delta$ and $\Psi$ parameters with Fe concentration is evidenced.

The amount of Fe doping in $ZrO_2$ films was controlled by varying the *m*:*n* = Fe:Zr binary process pulse ratio. We found that by changing the Fe:Zr ratio from Fe:Zr = 1:200 to Fe:Zr = 1:80 the amount of Fe increases from ~11% at. to ~19% at., as measured by XPS (Figure 3). Incidentally, we found that also SE measurements are sensitive to Fe doping (Figure 4). The SE spectra were analyzed by a multilayer model with a four-phase layered structure (substrate/film/surface rough layer/air).The complex refractive index of ALD Fe-doped $ZrO_2$ films was modeled using the Bruggeman effective medium approximation [18] with the two components $Fe_2O_3$ and $ZrO_2$. It turns out that the ellipsometric data can be sufficiently well fitted increasing the $Fe_2O_3$ volume fraction in $ZrO_2$ as the Fe:Zr pulse ratio increases. The XPS measurements give also us indications on the chemical state of Fe atoms. The peak energy and the shape profile, two peaks $Fe2p_{3/2}$ at 711.2 eV and $Fe2p_{1/2}$ at 724.7 eV split by spin orbit (intensity ratio 0.5) and a satellite centered at ~719.9 eV (Figure 3), is equivalent to the XPS profile seen for $Fe_2O_3$ [19]. It follows that Fe atoms are mainly in $Fe^{3+}$ chemical state in our $ZrO_2$ doped films. This finding is of particular relevance in the view of a potential development of ferromagnetism, because $Fe^{3+}$ maximizes the magnetization per atom.

The insertion of a 3+ cation in the $ZrO_2$ lattice favors the formation of oxygen vacancies, as reported from density functional theory (DFT) simulations of $Y^{3+}$ doped $ZrO_2$ [20]. Recent DFT simulations show that the same is true for Fe doping; at a Fe concentration equal or higher than 12% at. the stabilization of the cubic crystalline phase of $ZrO_2$ is favored [21].

To verify the crystallinity of our films, we performed XRD measurements on pure and Fe doped $ZrO_2$. Figure 5 reports XRD patterns of representative pure and Fe doped $ZrO_2$ before and after annealing at 600°C or 800°C, along with the reference diffraction patterns of tetragonal and monoclinic $ZrO_2$ phase [22]. When considering $ZrO_2$ films (Figure 5, top panel) the as grown film is crystalline and progressively develops the monoclinic phase after annealing; in the film exposed to 800°C, the monoclinic peak at ~32° is clearly visible. It should be noted that a monoclinic component already exists in the film annealed at 600°C, visible as a shoulder of the main tetragonal (or cubic) phase. The XRD pattern looks different in $ZrO_2$:Fe films (Figure 5, bottom panel), in particular the as grown film retains an amorphous character and undergoes crystallization in the tetragonal (or cubic) $ZrO_2$ phase when annealed at 600°C, where no evidence of $ZrO_2$ monoclinic phase appears. Only after annealing at 800°C a small monoclinic component can be seen. Thus, it appears that Fe doping is effective in inhibiting films crystallization during the growth process and in promoting the stabilization of the tetragonal (or cubic) phase in $ZrO_2$:Fe films, in agreement with theoretical findings. It should be mentioned that it is not possible, with conventional XRD, unambiguously discriminate between tetragonal and cubic ZrO$_2$ phase, because of the proximity of the related peaks; however DFT simulations evidences that cubic phase is energetically favored to form above few percent of Fe atomic content (data not shown).

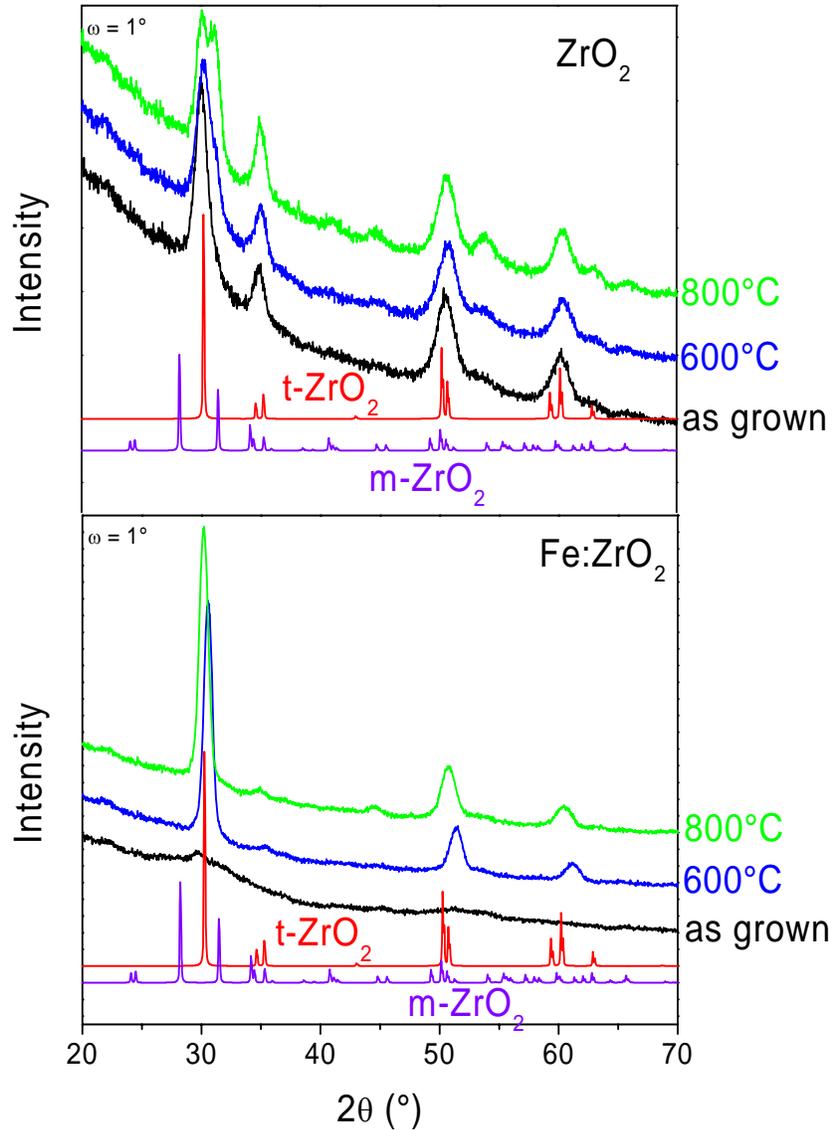

**Figure 5** – XRD measurements of as grown and annealed at 600°C or 800°C ZrO$_2$ (top) or ZrO$_2$:Fe (bottom) films. XRD reference patterns from t-ZrO$_2$ and m-ZrO$_2$ are also shown [22].

Further, from XRD measurements no evidence of Fe phase segregation is visible; however it would be possible that iron clustering exists in amorphous phase. ToF-SIMS depth profiles acquired on as grown and annealed (at 600°C) films, as shown in Figure 6(a) and (b) respectively, evidence a flat, uniform Fe profile along its entire thickness, without local intensity change, which indicates no Fe preferential aggregation. ToF-SIMS analysis also evidences a sharp and thermally stable interface between the films and the substrate with no Si diffusion, even after annealing; we can thus exclude any additional effects from Si unwanted doping, due to diffusion, on the $ZrO_2$:Fe phase stabilization. On the contrary, when considering ToF-SIMS depth profiles of films annealed at 800°C (not shown), Si is seen to diffuse in the film, affecting the interface with the substrate and, possibly, the film properties.

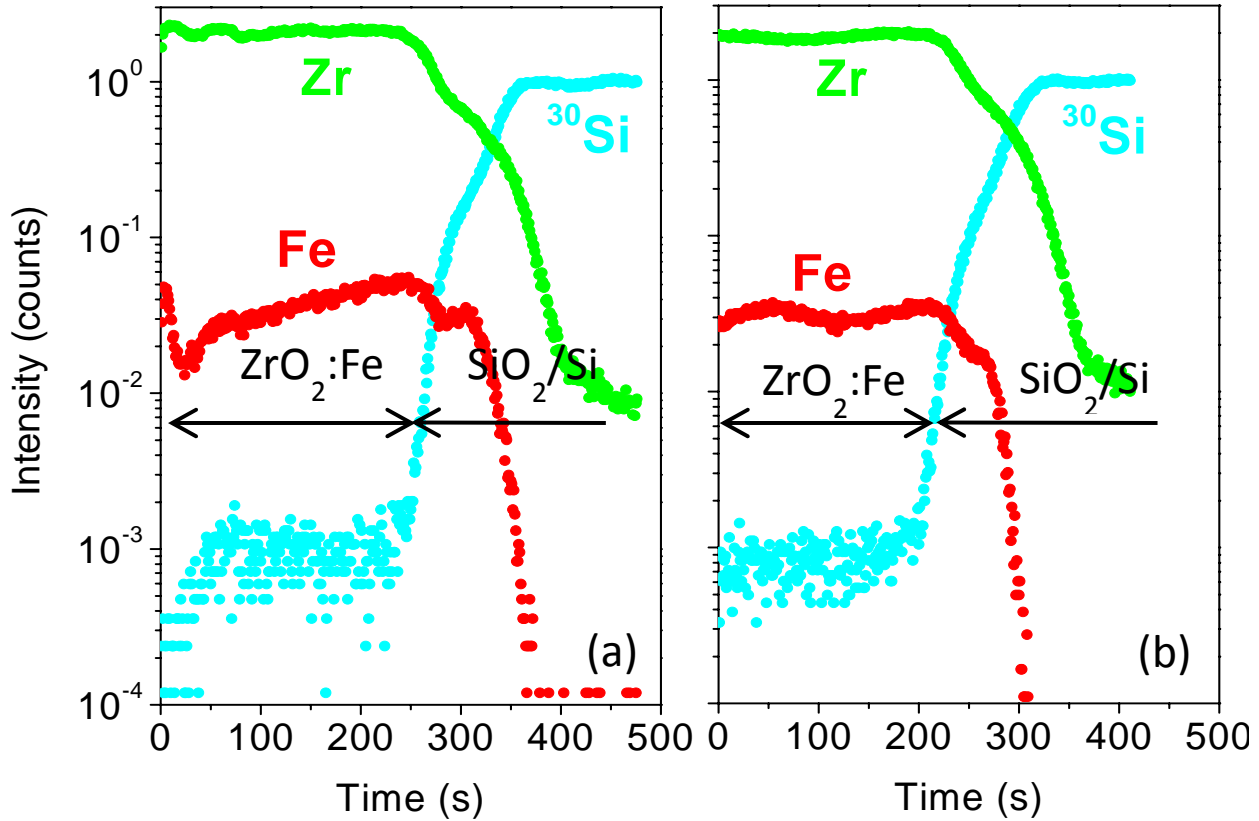

**Figure 6** – ToF-SIMS depth profiles of $ZrO_2$:Fe films; (a) as grown; (b) after annealing at 600°C.

A further confirmation of the controlled growth and thermal stability of pure and Fe doped $ZrO_2$ films comes out from XRR measurements and data fitting, as seen in Figure 7(a) and (b) for as grown and annealed at 600°C films. Data fitting, summarized in Table I, shows that for 18-19 nm thick films, the roughness is limited to 1.2 nm for pure $ZrO_2$ films and further reduced to 0.5 nm in $ZrO_2$:Fe, remaining unaltered after the annealing treatment. Further, the electron density indicates that Fe doping is effective in increasing film density with respect to pure $ZrO_2$.

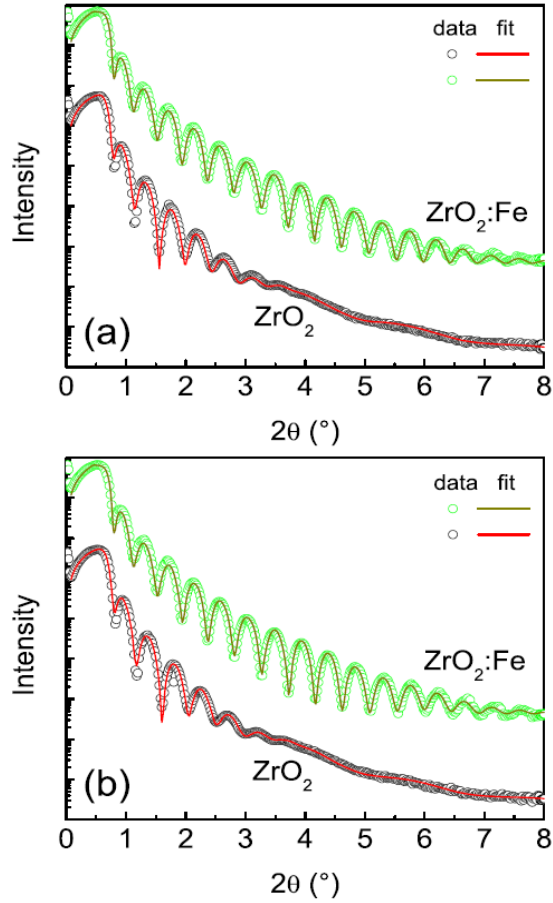

**Figure 7** – XRR measurements and fitting of $ZrO_2$ and $ZrO_2$:Fe films. (a) as grown; (b) after annealing at 600°C.

| **As grown** | Thickness $t$ (nm) | Roughness $\sigma$ (nm) | El. density $\rho_e$ (e Å$^{-3}$) |
|---|---|---|---|
| $ZrO_2$ | 18.6 | 1.1 | 1.53 |
| $ZrO_2$:Fe | 19.3 | 0.4 | 1.65 |
| **Annealed 600°C** | | | |
| $ZrO_2$ | 18.0 | 1.2 | 1.55 |
| $ZrO_2$:Fe | 19.2 | 0.5 | 1.64 |

**Table I** – XRR data fitting of $ZrO_2$ and $ZrO_2$:Fe films as grown and after annealing at 600°C.

## 4. Conclusions

We successfully developed an ALD process for the growth of Fe doped $ZrO_2$ films, where is possible to modulate the concentration of Fe in the films by varying the binary pulse ratio. Fe doping is effective in promoting the stabilization of tetragonal $ZrO_2$ crystalline phase after annealing at 600°C in $N_2$ flux for 60s, without formation of Fe clustering. Fe chemical state is confirmed to be $Fe^{3+}$ as seen by XPS, which maximizes the magnetic moments per atom. Further, the inclusion of Fe also favors the formation of oxygen vacancies, which seems to have a critical role in inferring the ferromagnetism in transition metal oxides, as predicted by theoretical simulations.

**Acknowledgments**

This work was partially supported by the project OSEA (Grant n.2009-2552) funded by Fondazione Cariplo.